\documentclass[conference]{IEEEtran}
\IEEEoverridecommandlockouts
\usepackage{indentfirst}
\usepackage{comment}
\usepackage{cite}
\usepackage{amsmath,amssymb,amsfonts}
\usepackage{algorithmic}
\usepackage{graphicx}
\usepackage{textcomp}
\usepackage[table]{xcolor}
\usepackage{subcaption}
\usepackage{forest}
\usepackage{dblfloatfix}
\usepackage{arydshln} 
\usepackage{multirow}
\usepackage{subcaption}  
\usepackage{musixtex}
\def\BibTeX{{\rm B\kern-.05em{\sc i\kern-.025em b}\kern-.08em
    T\kern-.1667em\lower.7ex\hbox{E}\kern-.125emX}}

\begin{document}
\renewcommand{\arraystretch}{1.3} 

\title{Relationships between Keywords and Strong Beats in Lyrical Music}

\author{\IEEEauthorblockN{Callie C. Liao}
\IEEEauthorblockA{\textit{Intellisky} \\
McLean, USA \\
ccliao@intellisky.org}
\and
\IEEEauthorblockN{Duoduo Liao}
\IEEEauthorblockA{\textit{School of Computing} \\
\textit{George Mason University}\\
Fairfax, USA \\
dliao2@gmu.edu}
\and
\IEEEauthorblockN{Ellie L. Zhang}
\IEEEauthorblockA{\textit{Intellisky} \\
McLean, USA \\
elzhang@intellisky.org}}

\maketitle

\begin{abstract}
Artificial Intelligence (AI) song generation has emerged as a popular topic, yet the focus on exploring the latent correlations between specific lyrical and rhythmic features remains limited. In contrast, this pilot study particularly investigates the relationships between keywords and rhythmically stressed features such as strong beats in songs. It focuses on several key elements: keywords or non-keywords, stressed or unstressed syllables, and strong or weak beats, with the aim of uncovering insightful correlations. Experimental results indicate that, on average, 80.8\% of keywords land on strong beats, whereas 62\% of non-keywords fall on weak beats. The relationship between stressed syllables and strong or weak beats is weak, revealing that keywords have the strongest relationships with strong beats. Additionally, the lyrics-rhythm matching score, a key matching metric measuring keywords on strong beats and non-keywords on weak beats across various time signatures, is 0.765, while the matching score for syllable types is 0.495. This study demonstrates that word types strongly align with their corresponding beat types, as evidenced by the distinct patterns, whereas syllable types exhibit a much weaker alignment. This disparity underscores the greater reliability of word types in capturing rhythmic structures in music, highlighting their crucial role in effective rhythmic matching and analysis. We also conclude that keywords that consistently align with strong beats are more reliable indicators of lyrics-rhythm associations, providing valuable insights for AI-driven song generation through enhanced structural analysis. Furthermore, our development of tailored Lyrics-Rhythm Matching (LRM) metrics maximizes lyrical alignments with corresponding beat stresses, and our novel LRM file format captures critical lyrical and rhythmic information without needing original sheet music.  
\end{abstract}


\section{Introduction}
With Artificial Intelligence (AI) song generation on the rise, developing algorithms capable of producing music that mimics the style of a human composer has become a significant and challenging area of research. When analyzing song lyrics, Linguistic Lyrical Elements (LLEs)—including keywords and non-keywords, stressed syllables and unstressed syllables, and other linguistic features—interact with musical structures such as rhythm and melody, playing an important role in shaping the overall music composition \cite{patel2008music}. In particular, lyrical keywords and rhythmically stressed syllables are critical for establishing musical interpretation and understanding. However, rhythmically stressed syllables may differ from dictionary-defined stressed syllables, and stress does not exist in all languages, such as character-based languages \cite{lerdahl1983generative} \cite{Temperley2001}. As a result, lyrical keyword extraction is an especially essential component of lyrical music composition. Keywords not only capture a broader range of literary expression but also help define the style of a song, guiding composers in conveying its central themes.  

The music establishes a rhythm and pitch structure consisting of sections, phrases, motifs, and other features \cite{TonalHarmony}. The rhythmic patterns are particularly related to the lyrics\cite{Temperley2001}, as songwriters often need to find a set of lyrics whose syllables and rhyming scheme would fit the desired rhythmic patterns created by the music, or vice versa. Thus, there is a strong association between lyrics and musical rhythmic patterns. 

Due to the lyrical keyword association with musical rhythmic patterns, rhythm and stress within rhythmic patterns play an equally critical role. Rhythm organizes “musical events in time” \cite{Latham2011} through rhythmically metrical stress, commonly categorized as downbeats, strong beats, and weak beats within a measure. Beats with metrical stress are \textit{strong beats}, whereas beats without metrical stress are \textit{weak beats}. Strong beats are essential for driving the rhythmic structure and emotional energy of a musical piece. While their placement varies with the time signature, strong beats consistently organize the rhythmic flow and enhance the music's emotional impact \cite{patel2008music}\cite{TonalHarmony}. Downbeats are a type of strong beat that specifically indicates the beginning of a measure. Since lyrics and rhythm are intricately connected, studying the correlation between each form of metrical stress can establish the fundamentals for AI music research. As a result, our investigation into the correlation between lyrical keywords and strong beats within musical scores can discover further insights.

Our key contributions are as follows:
\begin{itemize}
    \item We identified a strong correlation between lyrical keywords and strong beats, demonstrating that word types exhibit a significant overall tendency to align with corresponding beats compared to syllable types. These findings suggest that word type plays a more critical role in the effectiveness of musical analysis and the structure clarity of AI song generation, as connecting keywords with strong beats helps determine the placement of other lyrical and rhythmic elements.
    \item Our analysis of the identification of LLEs and their respective relationships further supports this strong correlation. These insights highlight that word types are more effective in establishing consistent associations within the musical framework than syllable types, particularly in ensuring rhythmic matching across various musical contexts.
    \item We developed tailored Lyrics-Rhythm Matching (LRM) metrics for analyzing the matching of beat stress and lyrical types, which can be applied to AI music generation research evaluations. The robust metrics are especially useful for naturally imbalanced data distributions, and they maximize stressed and unstressed LLE alignment with the corresponding strong and weak beats.
    \item Our development of the novel LRM file format effectively encapsulates critical lyrical and rhythmic information without the need for the original sheet music, enabling further analysis in lyrical music.
\end{itemize}
   
\section{Related Work}
There are statements surrounding the relationships between lyrically stressed features (i.e., keywords and stressed syllables) and rhythmically stressed features (i.e., downbeats and strong beats). For instance, the Linguistic Stress Rule states that strong beat alignment with stressed syllables is preferred by composers \cite{Temperley2022} \cite{Temperley2001}. Nevertheless, little quantitative research can be found, and much of the available past work mostly focuses on stressed syllables \cite{TemperleyTemperley2013} \cite{Bao2018NeuralMC} \cite{Zhang2022ReLyMe}.  

\cite{Nichols2009} presents an observational study of English popular music based on various correlations between categorical variables such as syllable stress, metric position, and stopwords \cite{Fox1989}. It also helps provide a quantitative correlation between lyrics and melodies. However, this paper seeks a general association between rhythmically accented notes and stressed syllables, emphasizing stressed syllables, which would result in a more limited range of applications. 

\cite{PalmerKelly1992} is a study focusing on relationships between linguistic prosody and musical meter, particularly noting the closer relationships of compound words and nuclear stress with musical metrical accent, as well as other noun and adjective relationships with musical accent. The conclusions in the study were based on vocalists' performances and the durations of nouns, adjectives, and syllables. It has a greater focus towards parts of speech and its impact on song performance, differing from our focus on studying specific lyrical and beat type relationships without a reliance on performance. 

Additionally, recent research reveals that stressed syllables are not the most fitting for musical alignment, as keywords and stressed beats have the closest relationship. \cite{Liao2022} proposes a multimodal lyrics-rhythm matching approach for audio that reveals a high probability of keywords landing on strong beats, which also contains limitations. The paper states that rhythmic or lyrical data were collected through the audio using beat tracking, which may result in potential misalignment errors or incorrect categorization. On the contrary, our approach uses information directly extracted from musical scores, guaranteeing the accuracy of our rhythmic and lyrical features.  

\section{Methodologies}
\subsection{Definitions}
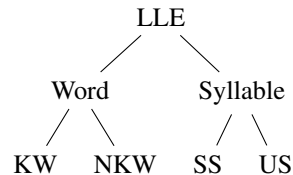
\begin{figure}[h]
    \centering
    \begin{forest}
  [LLE
    [Word
     [KW]
     [NKW]
     ]
    [Syllable
      [SS]
      [US]
    ]
]
\end{forest}
    \caption{The hierarchical structure of LLEs.}
    \label{fig:tree}
\end{figure}

The vocabulary defined for this research revolves around stress and lyrical types. The definitions are expanded in detail in Figure \ref{fig:tree} and Table \ref{table: type_Table}. KW, NKW, SS, US, SB, and WB represent keywords, non-keywords, stressed syllables, unstressed syllables, strong beats, and weak beats, respectively.

The hierarchy of the LLEs, previously denoted as Linguistic Lyrical Elements, is divided into lyrical and stress levels from top to bottom in Figure \ref{fig:tree}. The definition of LLEs is illustrated: LLE refers to the lyrical type of words and syllables indicated in the first layer; however, it represents the entirety of lyrically stressed and unstressed elements in the second layer. In Table \ref{table: type_Table}, the types in a song are divided into LLE and beat and categorized further by stress through an alternate perspective. The ``Type" represents the stress and element types present within a song. In particular, the elements under the beat type are strong and weak beats. The strong beats include downbeats---strong beats that land on the first beat of each measure---and other strong beats that do not land on the first beat, such as the second strong beat in 4/4 time.  

\begin{table}[h!]
\footnotesize
\centering
\begin{tabular}{|c|c|c|c| }
 \hline
 \multicolumn{2}{|c|}{Type}& Stressed & Unstressed \\
 \hline LLE  &  Word & KW & NKW \\
\cline{2-4}
      & Syllable & SS & US\\
\hline
Beat &  \cellcolor[HTML]{DDDDDD} & SB & WB \\
 \hline
 \end{tabular}
\caption{Major LLE and beat types present in a song, categorized by stress.}
\label{table: type_Table}
\end{table}

\subsection{The Dataset}
The dataset used for this research was collected from various piano books, resulting in the selection of 56 songs—26 in 3/4 time and 30 in 4/4 time. 

\subsubsection{LLE Counts}
The total word count is 2,713, with 3,516 syllables. Songs in 3/4 contain 1,123 words and 1,492 syllables, while those in 4/4 have 1,590 words and 2,024 syllables. Table \ref{table: count_table} shows the counts of different lyrical types in time signatures. ``All" refers to the total count of words or syllables across 3/4 and 4/4 time signatures. 

\begin{table}[h]
\footnotesize
\centering
\begin{tabular}{|p{1.2cm}|c c c|}

 \hline
 LLE & All &3/4&4/4\\
 \hline
Word    & 2,713& 1,123& 1,590\\
Syllable    & 3,516& 1,492& 2,024 \\
 \hline
 \end{tabular}
\caption{Counts of different lyrical types in different time signatures.}
\label{table: count_table}
\end{table}

\subsubsection{Lyrics-Rhythm Matching (LRM) Format}

For this pilot study, only relevant information is extracted from these songs. Each song is simplified into a format that aligns lyrics with their corresponding beats, rhythms, and notes, denoted as the Lyrics-Rhythm Matching (LRM) format. This format is specifically designed to analyze songs concerning lyrics and rhythm matching. Table \ref{table:LRM_file} shows an example of the LRM file for the folk song \textit{``Red River Valley"}.

In the LRM format, all lyrical and rhythmic information is collected from and accurately represented for each song without using the original sheet music.  Every phrase is separated by an asterisk ``*", each verse by a hashtag ``\#", and bridges are indicated by a percent ``\%".  For each song, the title, time signature, lyrics, rhythmic stress information, and duration for each syllable are recorded. 

Each lyrical word is followed by a group containing beat types and one or more pairs of parentheses. The number of groups corresponds to the number of syllables in the word. The number following the word represents the beat type: ``1" indicates a downbeat, ``2" denotes the second strong beat, and ``0" signifies a weak beat. Additionally, a ``-" before the number indicates a pickup note. Durations are recorded within the parentheses, which are expanded in more detail in Table \ref{table: note_type}.

\begin{table}[h]
\footnotesize

\begin{subtable}{0.4\linewidth}

\begin{tabular}{l} 
 \hline
 \textbf{TITLE:} Red River Valley\\
 \textbf{TIMESIG:} 4 4 \\
From -0 (4) \\
this -0 (4) \\
valley 1 (2) 2 (4)\\
they 0 (4)\\
say 1 (2)\\ 
you 2 (4)\\
are 0 (4)\\
going 1 (4) 0 (2.5,2) \\
* \\
We 0 (4) \\
shall 0 (4) \\
miss 1 (2) \\ 
your 2 (4) \\
bright 0 (4) \\
eyes 1 (2) \\
and 2 (4) \\
sweet 0 (4) \\
smile 1 (1,2) \\
*	  \\
\hline
 \end{tabular}
 \caption{A piece of the LRM file.}
\label{table:LRM_file}
    \end{subtable}
\hspace{0.1cm}
\begin{subtable} {0.6\linewidth}
 \begin{tabular}{cccl}
  \hline
 Pitch & Duration & Beat & Word\\
 \hline
C4 	 & 4 	&  0 	&  from 		 \\ 
C4 	&  4 	&  0 	&  this 		 \\
\hdashline
C4 & 	 2 & 	 1 	&  valley 	\\	 
C4 & 	 4 & 	 2 	  	& 	 \\
C4 & 	 4 & 	 0 & 	 they 	\\	 
\hdashline
C4 & 	 2 & 	 1 & 	 say 	\\	 
C4 	&  4 	&  2 	&  you 		\\ 
C4 	&  4 	&  0 & 	 are 	\\	 
\hdashline
C4 & 	 4&  	 1 & 	 going 	\\	 
C4 & 	 2.5&  	 0 	 &  	\\	 
\hdashline
C4 	&  2 	&  0 & 	 \\  		 
C4 	&  4 & 	 0 	&  we 	\\	 
C4 & 	 4 & 	 0 & 	 shall 	\\	 
\hdashline
C4 & 	 2 & 	 1 & 	 miss 	\\	 
C4 & 	 4 & 	 2 & 	 your 	\\	 
C4 & 	 4 & 	 0 & 	 bright \\		 
\hdashline
C4 	&  2 	&  1 & 	 eyes 	\\	 
C4 & 	 4 & 	 2 	&  and 	\\	 
C4 	&  4 & 	 0 	&  sweet 	\\	 
\hdashline
C4 & 	 1 	&  1 & 	 smile 	\\	 
 \hline
 \end{tabular}
  \caption{The derived music score.}
\label{table:score_table}
\end{subtable}
  \caption{Examples of a piece of the LRM file and corresponding derived music score information for the song  \textit{``Red River Valley"}.}
\end{table}

\begin{table}[h]
\footnotesize
\centering

\begin{tabular}{|>{\centering\arraybackslash}p{0.6cm}|>{\centering\arraybackslash}p{0.8cm}|>{\centering\arraybackslash}p{0.8cm}|>{\centering\arraybackslash}p{1cm}|>{\centering\arraybackslash}p{0.8cm}|>{\centering\arraybackslash}p{1.2cm}|}
 \hline
 Note & Whole & Half & Quarter & Eighth & Sixteenth\\
 \hline
 Code    & 1 & 2 & 4 & 8 & 16\\
 \hline
 Note & Dotted Whole & Dotted Half & Dotted Quarter & Dotted Eighth & Dotted Sixteenth\\
 \hline
 Code    & 1.5 & 2.5 & 4.5 & 8.5 & 16.5\\
 \hline
 \end{tabular}
\caption{Code representations of different musical note lengths.}
\label{table: note_type}
\end{table}

The LRM format only encodes the essential information for this research, as described above. Any additional information can be derived from the extracted song features.

\subsection{Extracted Musical Features}

As indicated in the LRM format, the extracted musical features from each song are only downbeats, strong beats, and weak beats. Other remaining non-pitch information, such as measures and beat placements, is derived from these available features through the code. Specifically, beat placements for downbeats, strong beats, and weak beats are derived from rhythmic metrical stress information and corresponding lyrical syllables. Once beat duration and placement information are obtained, the location of each measure is calculated. The derived score information can also help identify errors in beat duration inputs caused by human typographical mistakes or other reasons, further validating the LRM file.

Table \ref{table:score_table} shows the derived score from the LRM file of the song \textit{``Red River Valley"}. A row of dashes denotes one measure. In the first column, ``C4" is designated as the default pitch, as pitch information was not collected for this research. The second column presents the duration, the third column specifies the beat type, and the final column contains the corresponding word.

\subsection{Extracted Lyrical Features}

There are two types of LLEs extracted from the songs: word features, including KW and UKW, and syllable-type features, including SS and US. Details are as follows:

\subsubsection{Keyword Extraction}
The keywords are extracted using general keyword extraction methods, such as Rapid Automatic Keyword Extraction (RAKE) \cite{Rake2010} and Yet Another Keyword Extractor (YAKE)\cite{Yake2020}. The unextracted words are non-keywords. Table \ref{table: red_river_valley_kw_beats} shows an example of keywords and non-keywords extracted from the song \textit{``Red River Valley"}, along with their corresponding beat types. In this study, ``1" is used to indicate a keyword, while ``0" denotes a non-keyword.

\begin{table}[!htb]
    \begin{subtable}{.48\linewidth}
      \centering
        \begin{tabular}{lcc} 
 \hline
  Word & Keyword & Beat \\
 \hline
    from 	    &  0  & 0    	 \\ 
    this   	&   0 	&  0     		 \\
    valley 	& 1 & 	1   	 	\\	 
    they & 	  0 & 	  0 	\\	 
    say & 	 1  & 	1   	  	\\	 
    you 	&   0 &	 2 	\\ 
    are 	&  0 	&   0      	\\	 
    going & 	  1&  	1  	 	\\	
    we & 	 0  &  	 0      	\\	 
    shall 	&   0 	&  0    	 \\  		 
    miss 	&   1 & 	1    	\\	 
    your & 	0 & 	 2	  	\\	 
    bright & 	  1 &  0     		\\	 
    eyes & 	  1 &  1  	  	\\	 
    and & 	 0 & 	2	  \\		 
    sweet 	&   1 	& 0    	\\	 
    smile &  1 &  1  	\\	  
     &   &    	\\	  
     &   &    	\\	  
 \hline
        \end{tabular}
        \caption{Word Type}
 \label{table: red_river_valley_kw_beats}
    \end{subtable}
    \begin{subtable}{.48\linewidth}
      \centering
        \begin{tabular}{lcc}
             \hline
  Syllable & Stress & Beat  \\
 \hline
    from 	    & 1 & 0    	 \\ 
    this   	&  1 	& 0     		 \\
    val- 	& 1 & 	1   	 	\\	 
    ley   & 0 & 	2 	 \\
    they & 	 1 & 	 1 	\\	 
    say & 	 1  & 	1   	  	\\	 
    you 	&  1 &	2 	\\ 
    are 	&  1 	&  0      	\\	 
    go- & 	 1&  	1  	 	\\	
    ing & 	 0&  	 0     	 	\\
    we & 	 1 &  	0      	\\	 
    shall 	&  1 	&  0    	 \\  		 
    miss 	&  1 & 	1    	\\	 
    your & 	1 & 	2	  	\\	 
    bright & 	1 & 0     		\\	 
    eyes & 	 1 & 	1  	  	\\	 
    and & 	1 & 	2	  \\		 
    sweet 	&  1 	&  0    	\\	 
    smile & 1 & 	1  	\\	  
 \hline
        \end{tabular}
        \caption{Syllable Type}
 \label{table: red_river_valley_syll_beats}
    \end{subtable} 
    \caption{Examples of LLEs extracted from the song  \textit{``Red River Valley"}, along with their corresponding beat types.}
\end{table}

\subsubsection{Syllable Retrieval}
To retrieve the syllables from the lyrics, each lyrical word is split into one or more syllables using the Carnegie Mellon University (CMU) pronouncing dictionary\footnote{http://www.speech.cs.cmu.edu/cgi-bin/cmudict}. The CMU pronouncing dictionary produces a stress pattern for each respective word and marks the stressed syllables. ``1" indicates a stressed syllable, ``2" denotes the second stressed syllable, and ``0" signifies an unstressed syllable. In this research, we did not consider the second stressed syllable because it has relatively less importance than the first stressed syllable. Table \ref{table: red_river_valley_syll_beats} shows an example of stressed and unstressed syllables determined for the song \textit{``Red River Valley"}, along with their corresponding beat types.  

\subsubsection{LLE Distributions}
\begin{figure} [t]
  \centering
  \includegraphics[width=\columnwidth]{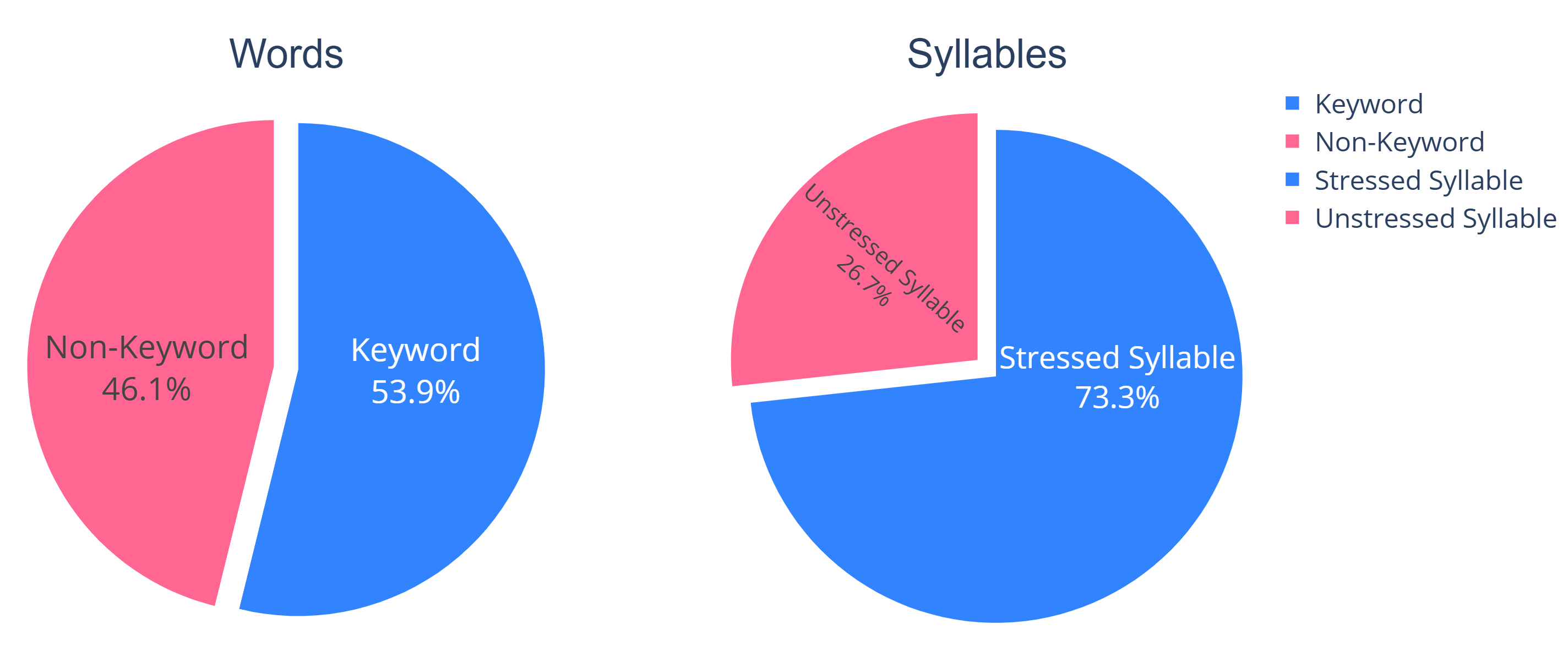}
  \caption{The overall percentages of lyrical features.}
  \label{fig:two_pie_word_syll}
\end{figure}

Figure \ref{fig:two_pie_word_syll} represents the distribution of overall percentages of the major lyrical features (i.e. LLEs) derived from the dataset. The figure illustrates that stressed syllables account for 73.3\% of the total syllable count, indicating that stressed syllables comprise nearly three-quarters of the lyrics overall. Using the YAKE \cite{Yake2020} keyword extraction method for this dataset, keywords constitute 53.9\% of the total word count, whereas non-keywords represent 46.1\%.

\subsection{Evaluation Metrics}
\subsubsection{Probability Alignment Evaluation}
Conditional probability is utilized to evaluate the alignment between lyrics and rhythm in this study. Conditional probability, denoted as $P(\textit{A}|\textit{B})$, calculates the probability of event $\textit{A}$ based on event $\textit{B}$\cite{Prob2010}. To assess the likelihood of LLEs aligning with beats of a specific type, let $\textit{A}$ represent a specific LLE, such as keywords, and $\textit{B}$ represent a beat type, such as strong beats. The conditional probability of LLEs landing on those beats is mathematically defined as:
\[
{\small
P(\textit{A}|\textit{B}) = \frac{P(\textit{A} \cap \textit{B})}{P(\textit{B})}
}
\]

where $P(\textit{A}|\textit{B})$ is the conditional probability of LLEs occurring given the presence of the beats, $P(\textit{A}\cap \textit{B})$ is the joint probability of both LLEs and beats occurring, and $P(\textit{B})$ is the probability of the beats occurring within the entire beat sequence.

\begin{figure*} [t]
    \centering
    \begin{subfigure}{0.48\textwidth}
      \includegraphics[width=1\textwidth]{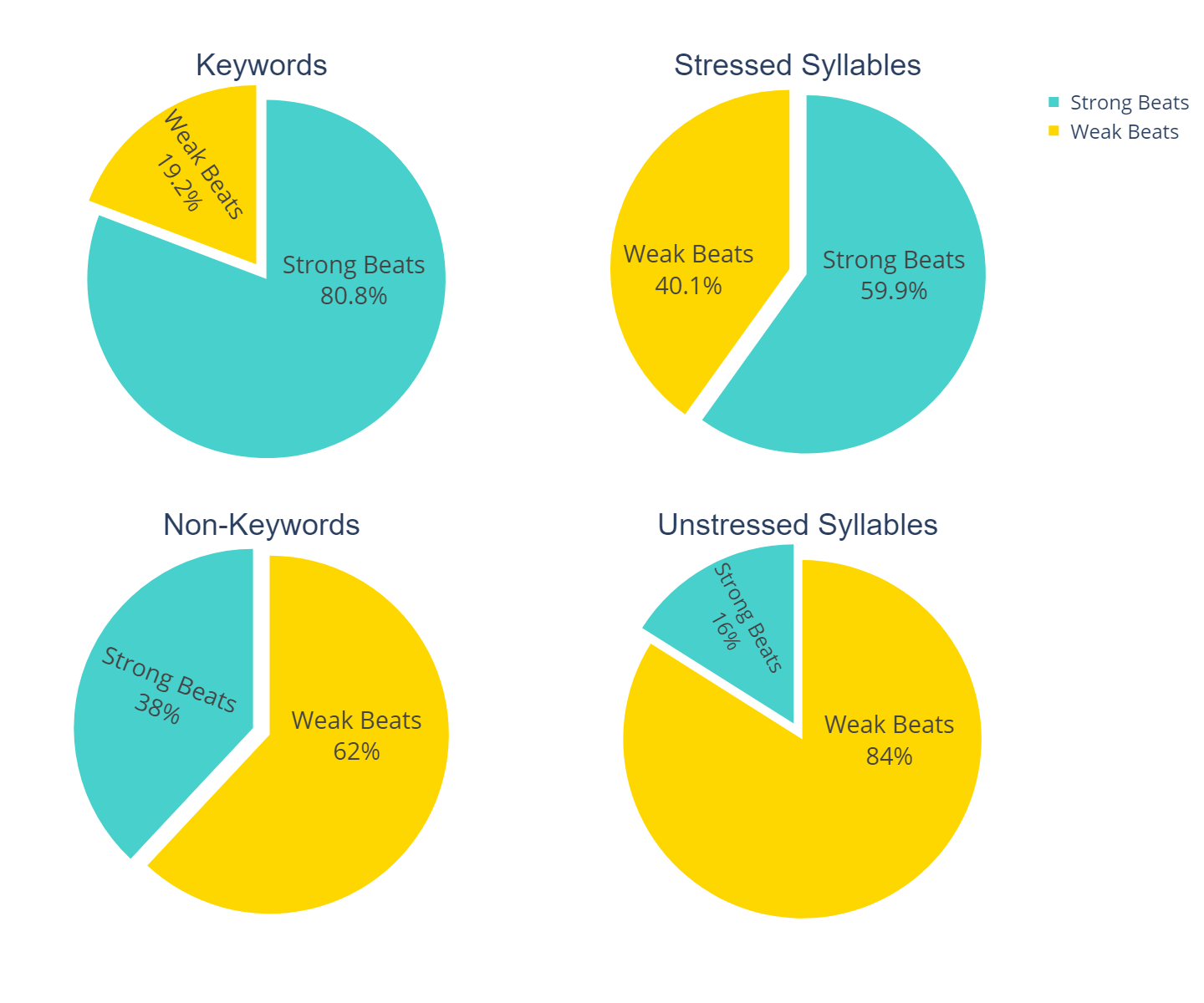}
      \caption{The overall percentages of strong and weak beats on lyrical LLE types.}
      \label{fig:four_pie_kw_ss_non-kw_un-ss}
    \end{subfigure}
     \hspace{1em}
    \begin{subfigure}{0.48\textwidth}
      \includegraphics[width=1\textwidth]{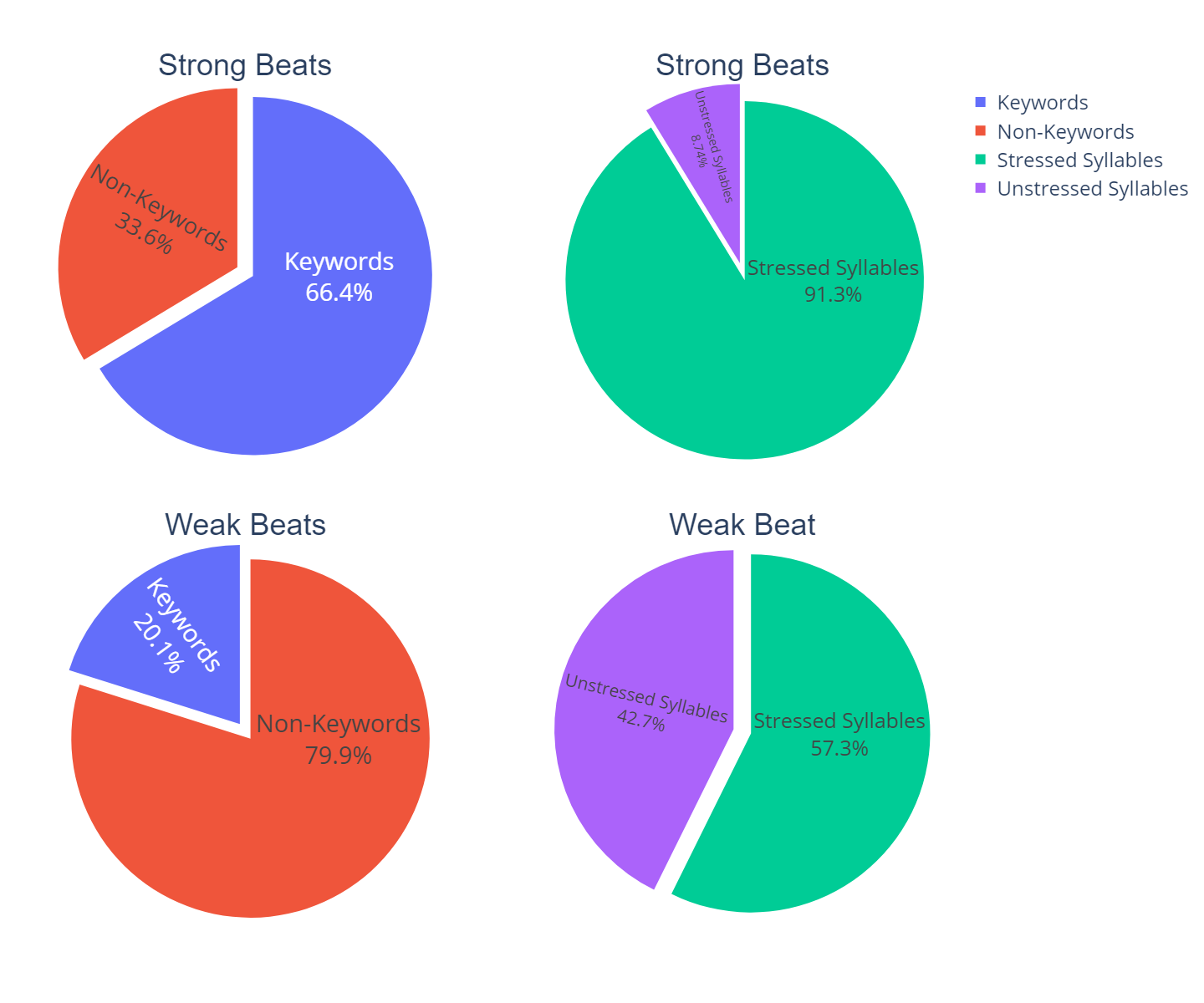}
      \caption{The overall percentages of lyrical LLE types on strong and weak beats.}
      \label{fig:four_pie_word_syll_sb_wb}
    \end{subfigure}
         \caption{The overall percentages of different cases between LLE types and beat types.}
     \label{fig:LRM_SCORE}
\end{figure*}

\subsubsection{Lyrics-Rhythm Matching Metrics}
The tendency of LLEs to align with specific beat types can be treated as a classification task. Specifically, stressed LLEs are expected to align with strong beats, while unstressed LLEs are predicted to land on weak beats. 

In this study, various classification metrics are utilized to assess the tendency of stressed LLEs to favor strong beats and unstressed LLEs to align with weak beats. These metrics are referred to as \textit{Lyrics-Rhythm Matching (LRM) Metrics}.

The \textit{Confusion Matrix} is a table used to evaluate the performance of a classification model by comparing the predicted labels with the actual labels \cite{Bishop2006}\cite{Powers2011}. Similarly, it can be customized as the \textit{LRM confusion matrix} to assess the matching of stressed LLEs (e.g., keywords or stressed syllables) to strong beats and unstressed LLEs (e.g., non-keywords or unstressed syllables) to weak beats as shown in Table \ref{table: LRM_conf_matrix}. 

\begin{table}[h]
\footnotesize
\centering
\begin{tabular}{r|c c}

  & Strong Beat &Weak Beat\\
 \hline
  Stressed LLE & SSB (TP)    & SWB (FN) \\
 Unstressed LLE & USB (FP)    & UWB (TN)\\
 \end{tabular}
\caption{LRM confusion matrix.}
\label{table: LRM_conf_matrix}
\end{table}

The \textit{LRM confusion matrix} summarizes the counts of Stressed LLEs on Strong Beats (SSB) or True Positive (TP), Unstressed LLEs on Strong Beats (USB) or False Positive (FP), Stressed LLEs on Weak Beats (SWB) or False Negative (FN), and Unstressed LLEs on Weak Beats (UWB) or True Negative (TN). It not only aids the understanding of the accuracy of the LRM, but also the types of mismatching it makes, providing deeper insights.  

From the \textit{LRM confusion matrix}, various performance metrics, \textit{LRM Metrics}, can be derived in the following.  

\textit{Accuracy} is the proportion of the correct matches (both SSB and USB) to the total number of instances. It is defined as:
{\small
\[
\textit{Accuracy} = \frac{\textit{SSB} + \textit{UWB}}{\textit{SSB} + \textit{SWB} + \textit{USB} + \textit{UWB}}
\]
}
\textit{Stress Matching (SM)} or \textit{Sensitivity} (or \textit{Recall}) is the proportion of stressed LLEs that are correctly falling on strong beats. It is defined as:
{\small
\[
\textit{SM} = \frac{\textit{SSB}}{\textit{SSB} + \textit{SWB}}
\]
}

\textit{Nonstress Matching (NM)} or \textit{Negative Predictive Value (NPV)} is the proportion of unstressed LLEs correctly landing on weak beats. It is defined as:
{\small
\[
\textit{NM} = \frac{\textit{UWB}}{\textit{SWB} + \textit{UWB}}
\]
}

 In the classification tasks, \textit{Sensitivity} and \textit{NPV} are both metrics that assess the performance of a classification model concerning different aspects of the data. Sensitivity focuses on detecting positives, while \textit{NPV} evaluates the correct classification of negatives. The harmonic mean of \textit{Sensitivity} and \textit{NPV} is employed to balance the performance in terms of both accurately detecting positive cases and minimizing false negatives in the negative class \cite{sokolova2009systematic}\cite{kwak2015evaluation}.
  
 Likewise, the score of Lyrics-Rhythm Matching (\textit{LRM Score or LRM-Score}), as the harmonic mean of \textit{SM} and \textit{NM}, serves as a robust evaluation metric for imbalanced data distributions, in particular to the scenarios with naturally imbalanced stressed and unstressed syllables. For instance, Figure \ref{fig:two_pie_word_syll} illustrates that approximately 75\% of the syllable types are stressed, while about 25\% are unstressed, indicating an imbalanced dataset of this study. 
 
  This metric can effectively maximize the alignment of stressed and unstressed LLEs with corresponding strong and weak beats by detecting stressed LLEs on strong beats and avoiding them on weak beats, as well as identifying unstressed LLEs on weak beats while minimizing them on strong beats. The formula of \textit{LRM Score} is defined as:
 {\small
\[
\textit{LRM-Score} =  \frac{2}{\frac{1}{\textit{SM}} + \frac{1}{\textit{NM}}}=  2\cdot\frac{{\textit{SM}}\cdot{\textit{NM}}}{{\textit{SM}} + \textit{NM}}
\]
}
 
\quad

\section{Experimental Results \& Analysis}

In this study, we explored potential correlations between the LLEs of lyrical features, particularly keywords, non-keywords, stressed syllables, and unstressed syllables, as well as the rhythmic features, including downbeats, strong beats, and weak beats. In the following experiments, keywords are extracted based on the YAKE \cite{Yake2020} method.

\subsection{The Comparison of LLE Distributions}
Figure \ref{fig:two_pie_word_syll} clearly indicates the distribution of overall percentages of the four LLEs derived from the dataset utilized in this experimental study. This distribution suggests that the proportion of keywords to non-keywords is approximately 1:1. However, stressed syllables comprise nearly three-quarters of the lyrics overall. The majority of syllables being stressed reveal that stressed syllables are not reliable indicators of the relationship between lyrics and rhythm, while the balanced distribution of keywords and non-keywords provides more accurate insights into these associations.

\subsection{Relationships between LLE and Beat Types}

Figure \ref{fig:four_pie_kw_ss_non-kw_un-ss} reveals the percentages of strong beats and weak beats that each lyrical feature lands on from the perspective of the overall lyrical features.  When comparing the percentage of strong beats across all four features, 80.8\% of keywords tend to land on strong beats the most, followed by stressed syllables with 59.9\%. This indicates that lyrical stress mostly aligns with rhythmic stress. However, keywords have a much clearer tendency to land on strong beats than stressed syllables, as the percentage of keywords landing on strong beats is 20.9\% higher than the percentage of stressed syllables landing on strong beats. Moreover, the stressed syllables were split into 40.1\% and 59.9\% for weak beats and strong beats respectively, which does not significantly present stressed syllables as having a stronger tendency to land on strong beats; instead, it presents stressed syllables as random in landing on a type of rhythmic feature. This further signifies that keywords are the most aligned with strong beats out of all the lyrical and rhythmic alignments. Conversely, for the percentage of weak beats, unstressed syllables have the greatest tendency to land on weak beats with an 84\% rate, followed by non-keywords with a 62\% rate. Thus, unstressed syllables are the most aligned with weak beats.

Figure \ref{fig:four_pie_word_syll_sb_wb} displays the distribution of each lyrical feature that rhythmic features are associated with across all the songs in the dataset. Stressed syllables predominantly align with strong beats, exhibiting a 91.3\% occurrence rate. However, they also appear on weak beats with a 57.3\% occurrence rate. According to Figure \ref{fig:two_pie_word_syll}, stressed syllables account for nearly three-quarters of all lyrical syllables. This high ratio leads to their predominance on strong beats, as their majority among all syllables limits the possibility for unstressed syllables to occupy strong beats. Nonetheless, because the percentage of weak beats occupied by stressed syllables is relatively close to the percentage of weak beats occupied by unstressed syllables, stressed syllables do not have a clear tendency to land on strong beats. In contrast, despite the nearly balanced distribution of keywords and non-keywords throughout the lyrics as illustrated in Figure \ref{fig:two_pie_word_syll}, keywords account for 66.4\% and 20.1\% of strong beats and weak beats in Figure \ref{fig:four_pie_word_syll_sb_wb}, respectively, displaying a clearer tendency to land on strong beats due to the disparity between the strong beat and weak beat percentages. Conversely, non-keywords are primarily associated with weak beats, occupying 79.9\% of the beats. The higher percentages of keywords on strong beats and non-keywords on weak beats indicate that these LLEs are strongly associated with their respective beat stresses, positioning keyword stress as a reliable indicator for lyrics-rhythm associations.

\begin{figure}[t]
\centering
  \includegraphics[width=0.98\columnwidth]{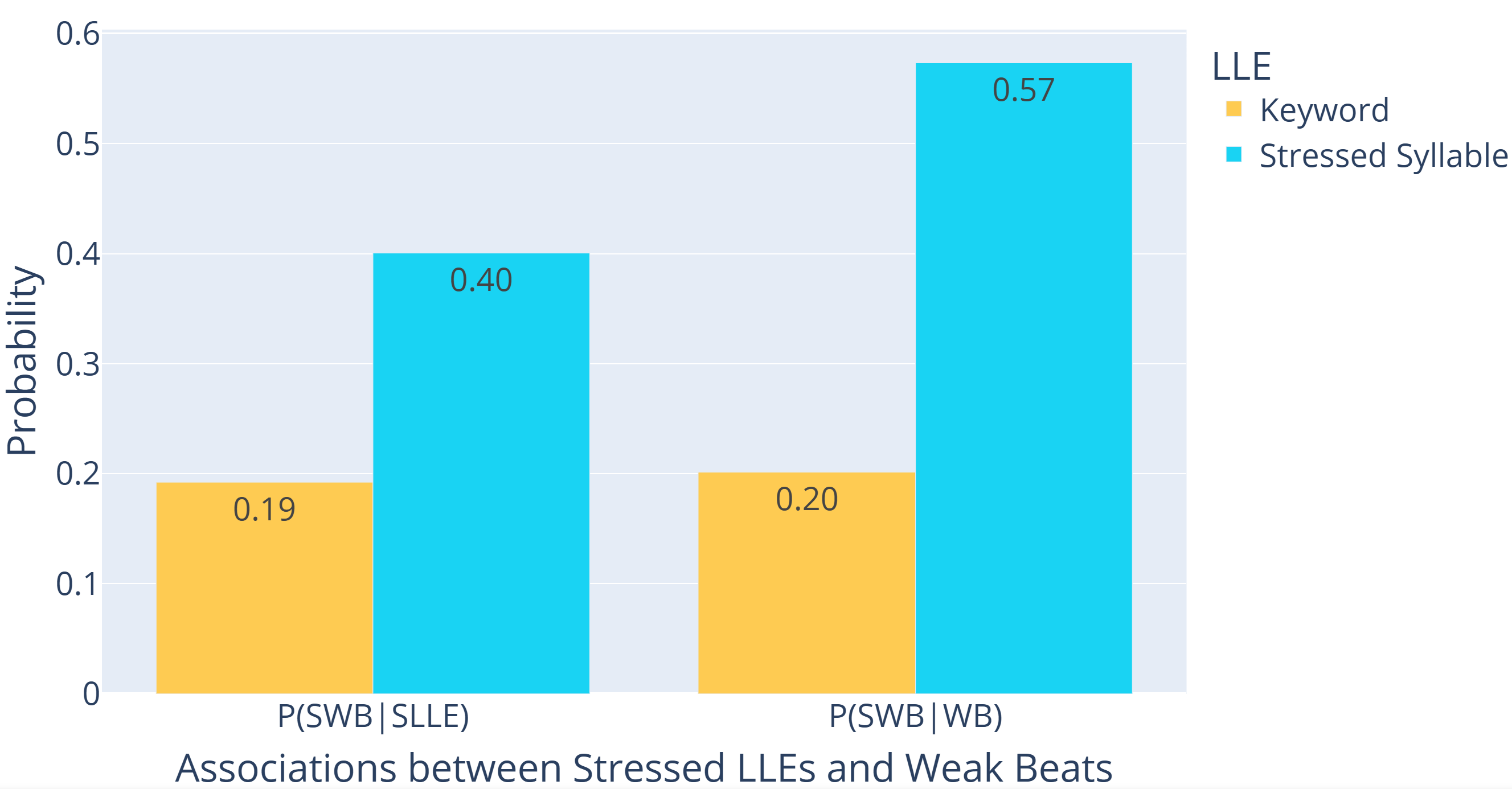}
  \vspace{0.2cm}
  \caption{Associations between stressed LLEs and weak beats.}
  \label{fig:two_bar_asso}
\end{figure}

\subsection{Associations between Stressed LLEs and Weak Beats}
\begin{figure}[t]
\centering
  \includegraphics[width=0.8\columnwidth]{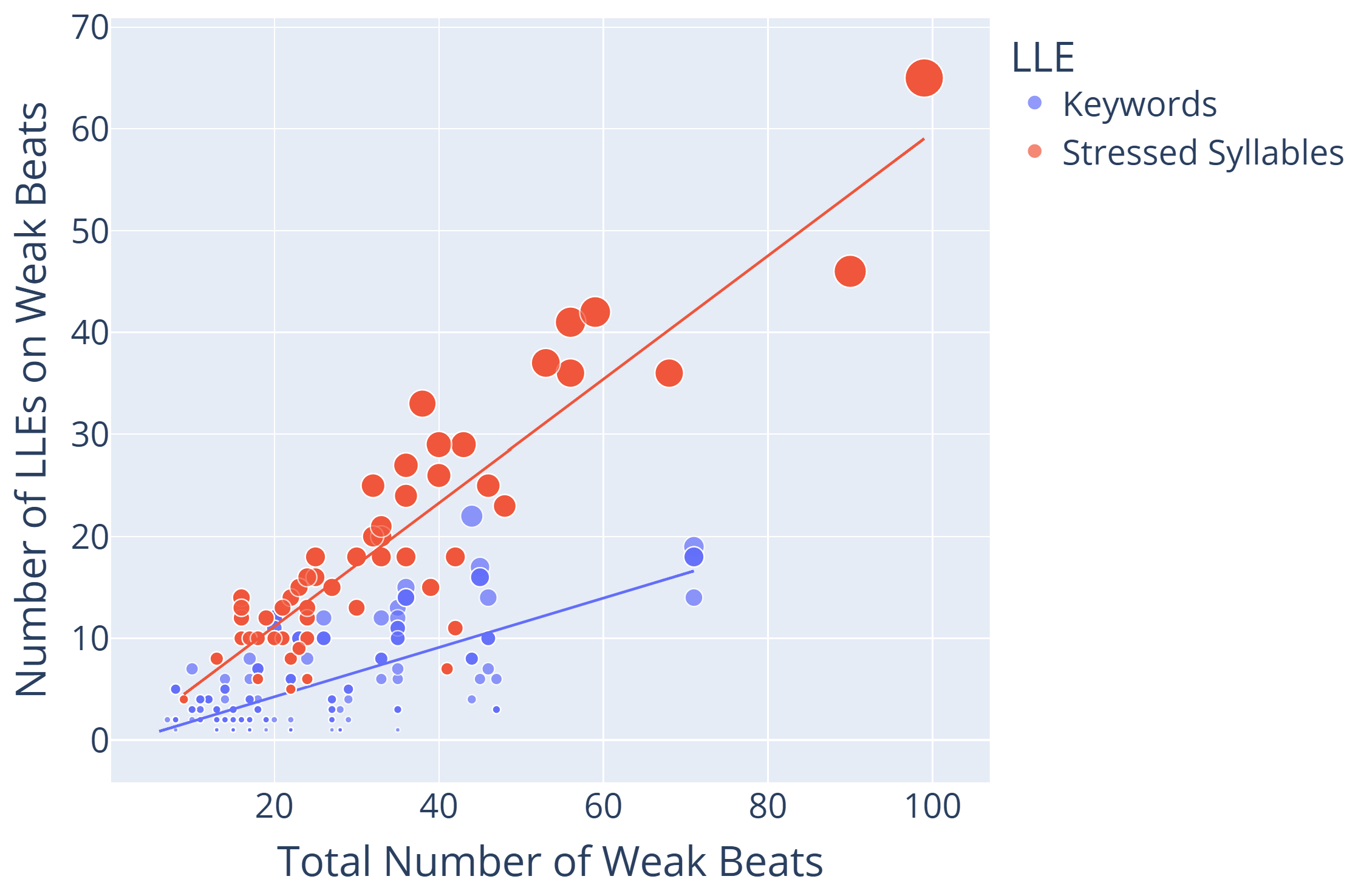}
  \caption{Relationships between the number of weak beats and the number of stressed LLEs on weak beats.}
  \label{fig:scatter_num_wb_type_wb}
\end{figure}

Figure \ref{fig:two_bar_asso} and Figure \ref{fig:scatter_num_wb_type_wb} demonstrate various comparative analyses to explore the associations between stressed LLEs and weak beats from multiple perspectives, including conditional probability and linear regression \cite{freedman2009statistical}.
\subsubsection{The Comparative Analysis of Conditional Probabilities}
Figure \ref{fig:two_bar_asso} presents a side-by-side comparison of the conditional probabilities for the associations between stressed LLEs and weak beats identified in Figure \ref{fig:LRM_SCORE}. The conditional probabilities are calculated based on all songs. The figure shows that the conditional probabilities of stressed syllables on weak beats, given stressed syllables and given weak beats, are $P(\textit{SWB}|\textit{SS})$ = 0.40 and $P(\textit{SWB}|\textit{WB})$ = 0.57, respectively, which demonstrates a degree of randomness and a slightly strong association between stressed syllables and weak beats. Conversely, the conditional probabilities of keywords falling on weak beats, given keywords and given weak beats, are $P(\textit{SWB}|\textit{KW})$ = 0.19 and $P(\textit{SWB}|\textit{WB})$ = 0.20, indicating that keywords clearly have a weak association with weak beats, as evidenced by consistently lower probabilities.

\subsubsection{The Comparative Analysis of Linear Regressions}

Figure \ref{fig:scatter_num_wb_type_wb} illustrates the relationship between the total number of weak beats and the number of LLEs on weak beats. The stressed syllables display a positive linear correlation, indicating that as the number of weak beats increases, the number of stressed syllables on weak beats also increases. Conversely, the keywords have a much slower increase, suggesting a weak positive correlation between the number of weak beats and the number of keywords landing on weak beats. As a result, the figure reveals that stressed syllables have a greater tendency to land on weak beats than keywords, indicating that stressed syllables do not align strongly with rhythmic stress tendencies in lyrics-rhythm associations.

\begin{figure}[t]
\centering
\includegraphics[width=0.8\columnwidth]{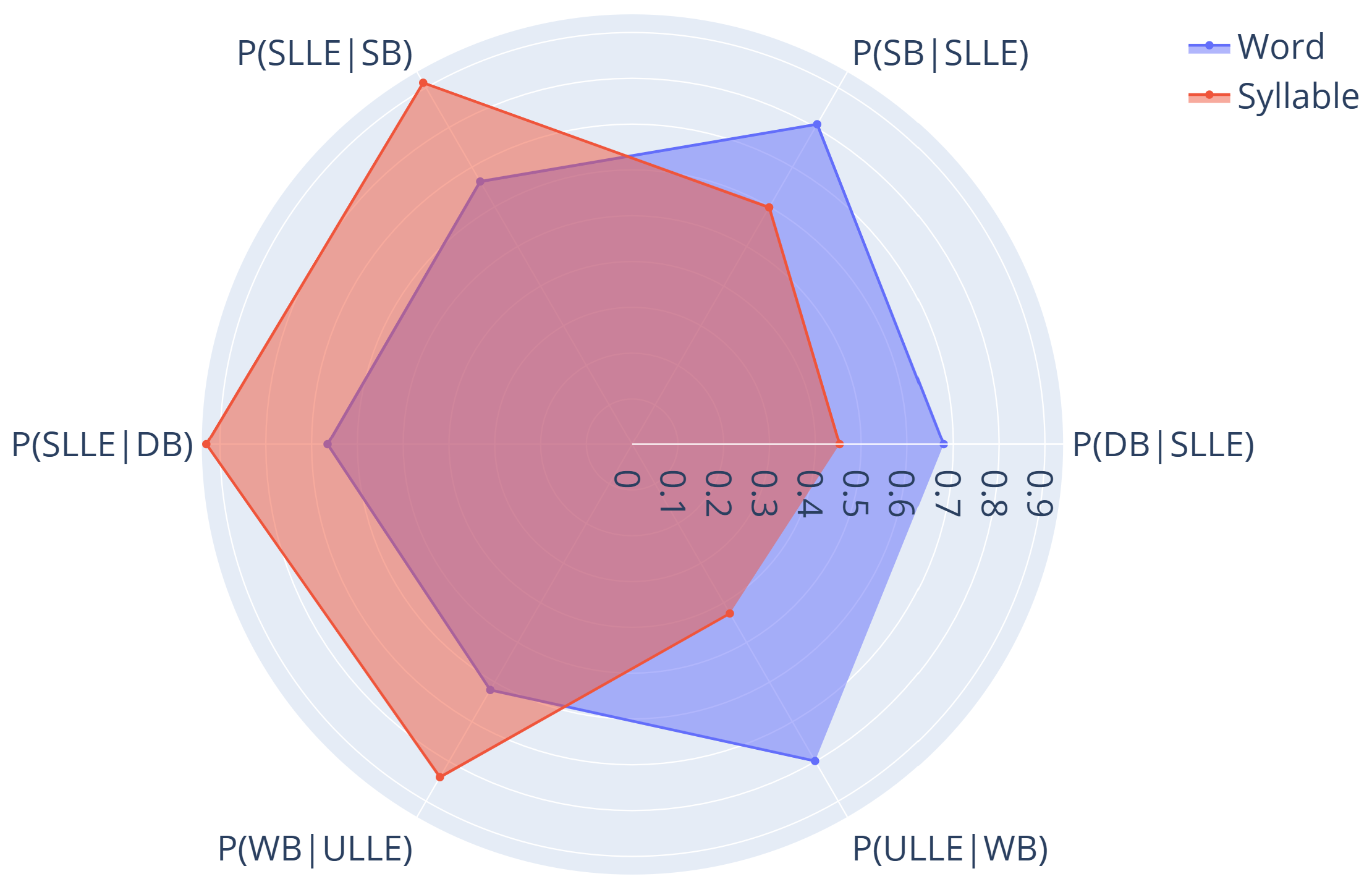}
  \caption{Comparisons for different conditional probabilities.}
  \label{fig:radar_chart}
\end{figure}

\begin{figure*}[h]
\centering
\includegraphics[width=0.9\textwidth]{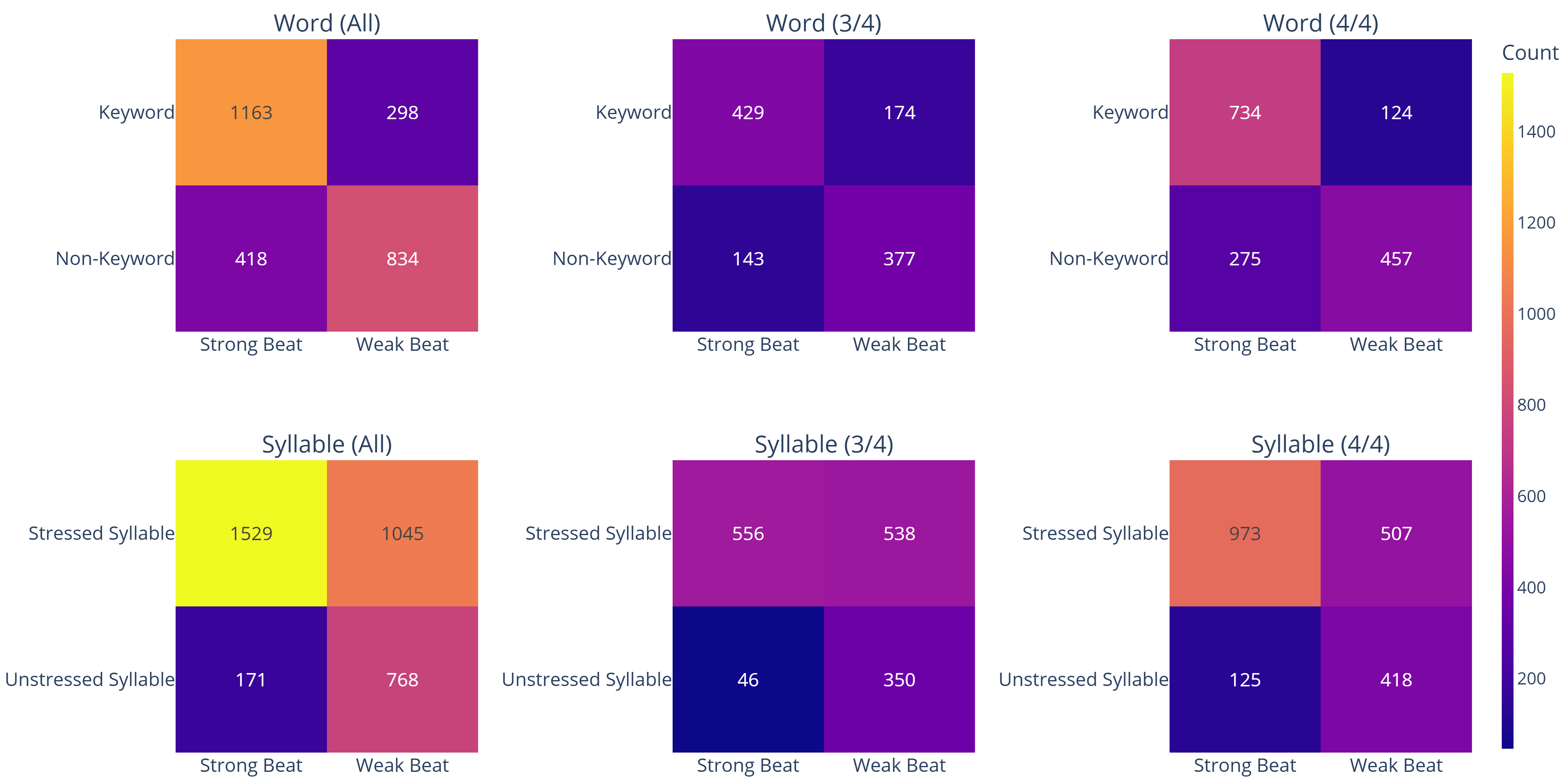}
  \caption{The confusion matrices for different cases.}
  \label{fig:confusion_matrix}
\end{figure*}

\subsection{Comparisons for Different Conditional Probabilities}

Comparisons of different conditional probabilities reveal the varying associations between LLEs and beats, offering insights into the overall relationships between lyric types and beats.

Figure \ref{fig:radar_chart} displays a radar chart of the relationships between lyric types (i.e., word and syllable) with various beats based on their conditional probabilities. In the figure, the symbols are defined as $\textit{DB}$ for downbeats, $\textit{SLLE}$ for stressed LLEs, and $\textit{ULLE}$ for unstressed LLEs. $P(\textit{SLLE}|\textit{SB})$ and $P(\textit{SLLE}|\textit{DB})$ represent the conditional probabilities of stressed LLEs given strong beats and downbeats, respectively. Similarly, $P(\textit{SB}|\textit{SLLE})$ and $P(\textit{DB}|\textit{SLLE})$ represent the conditional probabilities of strong beats and downbeats given stressed LLEs. $P(\textit{ULLE}|\textit{WB})$ and $P(\textit{WB}|\textit{ULLE})$ denote the conditional probabilities of unstressed LLEs given weak beats and weak beats given unstressed LLEs, respectively.

On the right side of the radar map, the word relationships display higher conditional probabilities for $P(\textit{SB}|\textit{KW}) \approx$ 0.8 and $P(\textit{DB}|\textit{KW}) \approx$ 0.7, indicating that a higher percentage of keywords land on strong beats and downbeats, respectively. On the contrary, $P(\textit{SB} \vert \textit{SS}) \approx$ 0.6 and $P(\textit{DB} \vert \textit{SS}) \approx$ 0.5, indicating that a lower percentage of stressed syllables land on strong beats and downbeats and suggesting that there is a greater percentage of stressed syllables landing on weak beats than keywords. For keywords, $P(\textit{NKW}|\textit{WB}) \approx$ 0.8, which is compared to the $P(\textit{WB}|\textit{USS}) \approx$ 0.4. This reveals a significant percentage of weak beats consisting of non-keywords and a small percentage of unstressed syllables consisting of weak beats. The higher alignment of more keywords landing on strong beats and downbeats, as well as the higher percentage of weak beats consisting of non-keywords suggests that keywords, the stressed word type, exhibit a strong direct relationship with strong beats, the stressed beat type. 

On the left side of the radar map, the syllable relationships display high conditional probabilities for $P(\textit{SS}|\textit{SB}) \approx$ 0.9, $P(\textit{SS}|\textit{DB}) \approx$ 0.9, and $P(\textit{WB}|\textit{US}) \approx$ 0.8. This demonstrates that a high percentage of strong beats and downbeats consist of stressed syllables, and that a high percentage of unstressed syllables consist of weak beats. However, when compared to the conditional probabilities calculated earlier on the opposite end, the disparities are approximately 0.3, 0.4, and 0.4, displaying a significant gap and indicating that there is an abundance of stressed syllables, leading to significant coverage of downbeats and strong beats, yet there is not a clear association within the stressed syllables since there is a high percentage of the remaining stressed syllables that do not land on strong beats. For keywords, the disparities between $P(\textit{SB}|\textit{KW})$ and $P(\textit{KW}|\textit{SB})$, $P(\textit{DB}|\textit{KW})$ and $P(\textit{KW}|\textit{DB})$, and $P(\textit{NKW}|\textit{WB})$ and $P(\textit{WB}|\textit{NKW})$ are approximately 0.1, 0, and 0.2. The disparities are much lesser than those for stressed syllables, indicating greater consistency among the number of keywords, strong beats, downbeats, and overall stressed word and beat types. Furthermore, since the percentages are relatively high in the conditional probabilities for keywords, it reinforces the strong positive correlation between keywords and strong beats. In the entirety of the radar map, the conditional probabilities used match with the formulas of the SM and NM metrics described earlier, hence reflecting the robustness of the LRM-Score.

 \subsection{Evaluation Using LRM Metrics}

Figure \ref{fig:confusion_matrix} illustrates the evaluations of two distinct LLE types across various time signatures, represented through confusion matrices. In the figure, The overall confusion matrices clearly indicate that word types exhibit a significant overall tendency to align with corresponding beats compared to syllable types, highlighting their effectiveness in rhythmic matching across various musical contexts.

\begin{table}[t]
\centering
\small
\hfill
\begin{tabular} {|c|c||cccc|}
 \hline
 Type& TS&Accuracy&SM&NM&LRM Score\\
 \hline
        & All & 0.736& 0.796& 0.737& 0.765\\
Word    & 3/4 & 0.718& 0.711& 0.684&  0.698\\
        & 4/4 & \cellcolor[HTML]{DDDDDD}0.749&  \cellcolor[HTML]{DDDDDD}0.855&  \cellcolor[HTML]{DDDDDD}0.787&  \cellcolor[HTML]{DDDDDD}0.820\\
 \hline
   & All & 0.654& 0.594& 0.424& 0.495\\
Syllable   & 3/4 & 0.608& 0.508& 0.394& 0.444\\
   & 4/4 & 0.688& 0.657& 0.452& 0.536\\
 \hline
\end{tabular}
\caption{Matching metrics for stressed or unstressed lyrical types aligning with corresponding beats across various time signatures.}
\label{table: metrics_table}
\end{table}

Table \ref{table: metrics_table} presents the LRM metrics—specifically accuracy, SM, NM, and theLRM score—for various lyrical types when analyzed across all time signatures and two specific time signatures, 3/4 and 4/4. In the 4/4 time signature, the word type demonstrates the highest accuracy (0.749), SM (0.855), and NM (0.787), while the syllable type exhibits lower accuracy (0.688), SM (0.657), and NM (0.452). Similarly, in the 3/4 time signature, word types again show higher accuracy (0.718), SM (0.711), and NM (0.684), whereas syllable types have the lowest accuracy ((0.608), SM (0.508), and NM (0.394).

These results suggest that word types are more reliable in their associations with beats in both time signatures, making them particularly valuable for musical analysis and composition. Their strong performance in accuracy and precision emphasizes their importance in understanding rhythmic structures, especially within the 4/4 time signature.

\begin{figure}[h]
 \centering
  \includegraphics[width=0.95\columnwidth]{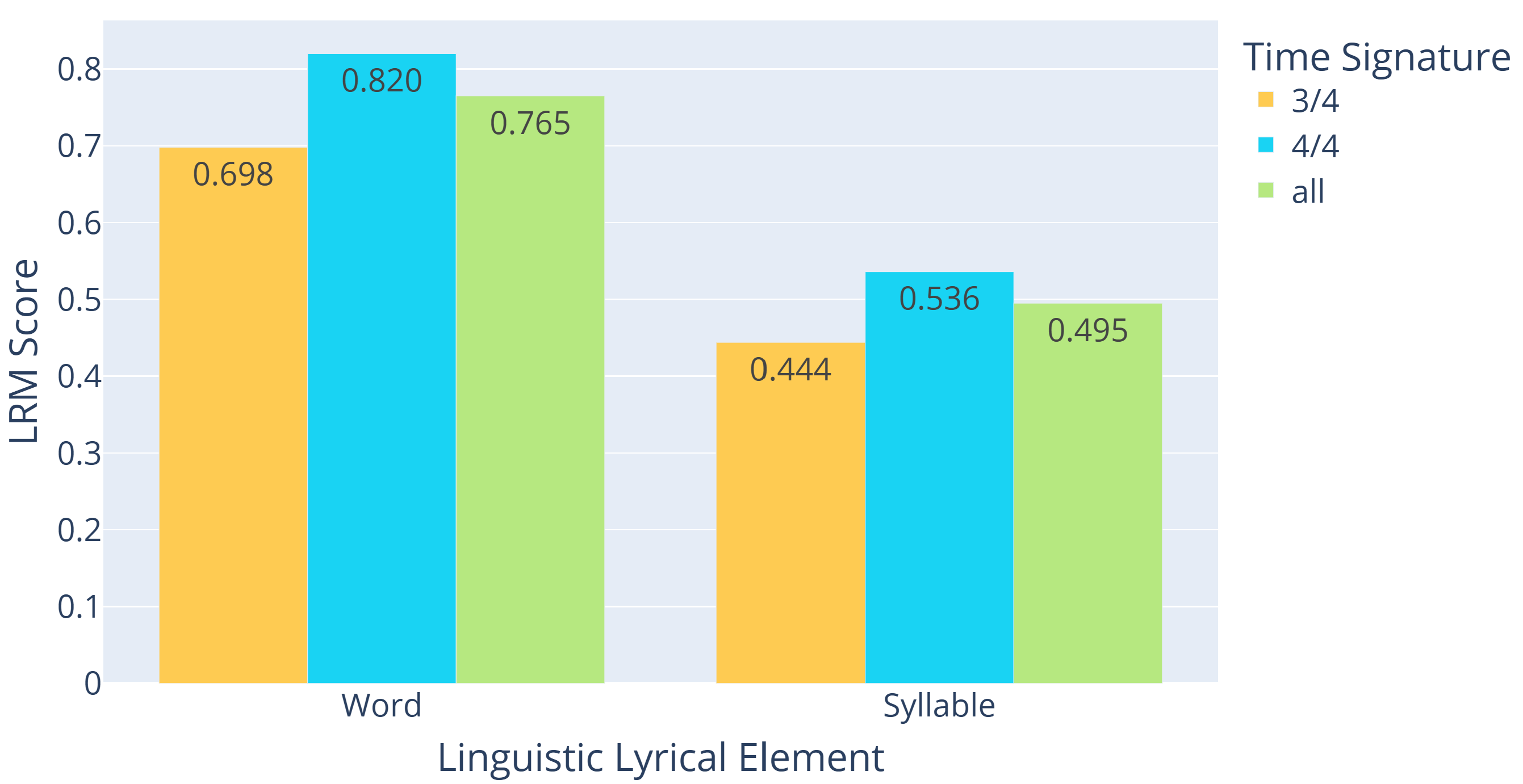}
  \caption{Matching scores for different lyrical types with different time signatures}
  \label{fig:rank_bar_avg}
\end{figure}

In Table \ref{table: metrics_table} and Figure \ref{fig:rank_bar_avg}, the LRM scores for word types across different time signatures are approximately 0.765, 0.698, and 0.820 for all, 3/4, and 4/4 time signatures, significantly exceeding the LRM scores for syllable types, which are 0.495, 0.444, and 0.536, respectively. It clearly demonstrates that word types exhibit a markedly stronger tendency to land on their corresponding beat types across all time signatures examined in this study, as evidenced by clear and notable patterns. In contrast, syllable types reveal a significantly weaker tendency to correspond with associated beat types. This disparity underscores the reliability of word types in capturing rhythmic structures in music, highlighting their essential role in effective rhythmic matching and analysis.

\section{Conclusions \& Future Work}
 
Aligning rhythmic features with lyrical features is critical for AI song generation, and focusing on lyrical stress and rhythmic stress can help maintain the momentum of a melody. This paper investigates the relationships between lyrically stressed features (i.e., keywords and stressed syllables) and rhythmically stressed features (i.e., downbeats and strong beats). It reveals correlations among unstressed rhythmic and lyrical features (i.e., weak beats, non-keywords, and unstressed syllables) to demonstrate that keywords not only tend to land on strong beats but also create the most closely aligned pattern with the original rhythmic beat pattern of each song. The experimental results include 76.8\% of the keywords landing on strong beats and 72\% of the non-keywords landing on weak beats. Moreover, on average, 83\% of the strong beats are comprised of keywords, leaving only 17\% with non-keywords. These findings, along with the identification of LLEs, strongly suggest that word types are more effective at establishing consistent associations within the musical framework compared to syllable types. Additionally, our development of the LRM metrics for beat stress and lyrical type analysis is robust and can maximize LLE alignment with various beat stresses, contributing to AI music generation research evaluations. Furthermore, our novel LRM file format extracts lyrical and rhythmic information in a simple form, aiding lyrical music analysis without the need for original sheet music.

Additionally, the LRM scores, a key LRM metric, report approximately 0.765 when measuring keywords on strong beats and non-keywords on weak beats across various time signatures and 0.495 for syllable types. This further highlights the stronger alignment of word types with beat types than syllable types, demonstrating their superior reliability in capturing rhythmic structures and supporting effective rhythmic analysis.

This pilot study is subject to certain limitations in terms of data scope. The dataset consists of only 56 songs, all of which feature 3/4 or 4/4 time signatures, resulting in a small and highly specific sample. Future work will aim to expand the dataset by including a larger and more diverse selection of songs from various musical genres, thereby enabling a more comprehensive analysis and further validation of our conclusions. Additionally, a broader range of keyword extraction methods will be incorporated to enhance the alignment between keywords and strong beats.

\bibliography{conference_kw_sb}
\bibliographystyle{IEEEtran}

\end{document}